\documentclass[preprint]{aastex}
\input psfig.sty

\newcommand{\nh}{N_{\rm H}}
\newcommand{\sax}{{\it Beppo\-SAX}}
\newcommand{\source}{XTE\ J1118+480}
\newcommand{\msun}{{\rm M}_\sun}
\newcommand{\xte}{{\it RXTE}}

\newcommand{\asm}{{\it RXTE}/ASM}
\newcommand{\chandra}{{\it Chandra}}

\newcommand{\ergcms}{\mbox{erg cm$^{-2}$ s$^{-1}$}}
\newcommand{\ergs}{\mbox{ erg s$^{-1}$}}

\begin{document}

\title{Spectral and temporal  behavior of  the black hole candidate
 XTE J1118+480 as observed with \sax} 

\author{F. Frontera\altaffilmark{1,2},
L. Amati\altaffilmark{2},
A. A.~Zdziarski\altaffilmark{3},
T. Belloni\altaffilmark{4},
S.~Del~Sordo\altaffilmark{5},
N. Masetti\altaffilmark{2},
M.~Orlandini\altaffilmark{2},
E. Palazzi\altaffilmark{2}
}
\altaffiltext{1}{Dipartimento di Fisica, Universit\`a degli Studi di Ferrara,
Via Paradiso 12, I-44100 Ferrara, Italy; frontera@fe.infn.it}

\altaffiltext{2}{Istituto Astrofisica Spaziale e Fisica Cosmica, CNR,
Via Gobetti 101, I-40129 Bologna, Italy; frontera@bo.iasf.cnr.it}

\altaffiltext{3}{N. Copernicus Astronomical Center, Bartycka 18,
00-716 Warsaw, Poland; aaz@camk.edu.pl}

\altaffiltext{4}{Osservatorio Astronomico di Brera, Via Bianchi, 46, I-23807
Merate, Italy}

\altaffiltext{5}{Istituto Astrofisica Spaziale e Fisica Cosmica, CNR,
Via U. La Malfa 153, I-90146 Palermo, Italy}

\begin{abstract}
XTE J1118+480 is a well established black hole candidate with a 
mass estimate in the range from 7 to 10 solar masses. With \sax\ 
we observed the source 4 times, from April to December 2000. Results of the first 
observation were already reported (Frontera et al. 2001). Here we report 
spectral results of the later observations, performed in May, June and 
December 2000 and compare them with the results obtained from the 2000 
April observation. We observe a decrease of the column density from a value
consistent with the Galactic $N_{\rm H}^{\rm G}$ obtained from radio measurements
to a value a factor 2 lower. The spectra are well fit with a thermal 
Comptonization plus a blackbody model. The blackbody luminosity 
decreases with time, while the electron temperature of the Comptonizing electrons
does not show significant changes. A Compton reflection 
component is apparent and stable, although weak (mean value of 
$\Omega/2\pi = 0.21^{+0.05}_{-0.04}$). The reflector shows
a low metallicity (mean value of $Z/Z_\odot = 0.13^{+0.06}_{-0.04}$). 
On the basis of the spectral results, a hot central disk appears the best scenario 
for the high energy photons, while the temporal properties point to a non thermal
origin of a fraction of the soft X--ray photons, likely synchrotron emission 
internal to the hot disk.

\end{abstract}

\keywords{accretion, accretion disks --- binaries: general ---  black hole
physics --- stars: individual (XTE J1118+480) --- X-rays: observations ---
X-rays: stars}

\section {Introduction}
\label{s:intro}

\begin{figure}[t!]
\epsscale{1.0}
\plotone{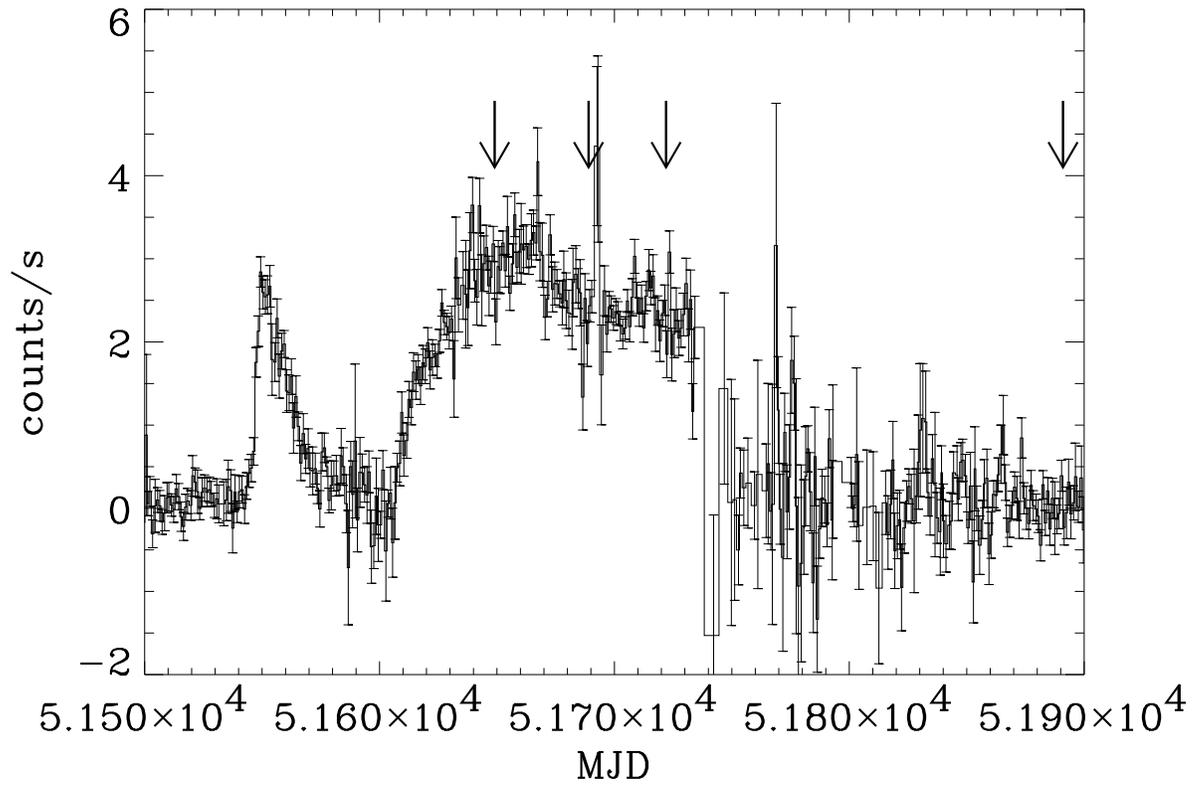}
\caption{One-day-average count rates from \source\/ detected
with the \asm\/ during the outburst. (From the Internet
public archive at xte.mit.edu/XTE/asmlc/ASM.html.)
The times of the \sax\/ observations are marked with the arrows. 
}
\label{f:outburst}
\end{figure}

The high Galactic latitude transient source \source\ was discovered
with the All-Sky Monitor (ASM) aboard
the {\it Rossi X-Ray Time Explorer\/} (\xte) satellite \cite{Remillard00}
during a low state mini--outburst in X--rays
(see Fig.~\ref{f:outburst}). The source has attracted the interest of many
investigators, who have performed observational and theoretical studies.
The radio, near-infrared, optical and EUV counterparts of \source\ were
discovered soon after the ASM announcement
(Pooley \& Waldram 2000, Chaty et al. 2000, Uemura et al. 2000a, Uemura et al. 2000c,
Mauche et al. 2000\nocite{Pooley00,Chaty00,Uemura00a,Uemura00c,Mauche00}) and 
their temporal and spectral properties reported (Cook et al. 2000, Garcia et al.
2000, Haswell, Hynes \& King 2000a, McClintock et al. 2000, Uemura et al. 2000c, Wagner
et al. 2000\nocite{Cook00,Garcia00,Haswell00a,McClintock00,Uemura00b,Uemura00c,Wagner00}). 
The Low Mass X--ray Binary (LXMB) nature of the system was soon established 
\cite{Uemura00c}, even though a surprisingly low X--ray--to--optical flux ratio 
of $\sim 5$, against a typical value of $\sim 500$ \cite{Jvp95}, was noticed 
\cite{Garcia00}. The binary period of the system ($P = 0.1703$~days) and its 
mass function  were also determined 
\cite{McClintock00,McClintock01a,Wagner00}.
The estimate of a large value of the mass funtion ($\sim 6\, M_\odot$,
McClintock et al. 2000, 2001a\nocite{McClintock00,McClintock01a}; Wagner
et al. 2000\nocite{Wagner00}) provided evidence
of the black hole nature of \source. Wagner et al. (2001)\nocite{Wagner01} 
provided the following values of the sytem parameters: inclination 
$i = 81 \pm 2$~degrees, black hole mass $M_{\rm X}/\msun = $6.0--7.7, distance
$d = 1.9\pm 0.4$~kpc. We also derived the parameters of the system 
(Frontera et al. 2001b, hereafter F01\nocite{Frontera01b}), finding values 
 which are consistent with those derived by 
McClintock et al. (2001a) and Wagner et al. (2001)\nocite{McClintock01a,Wagner01}.  

The multiwavelength spectrum of XTE J1118+480, from radio to X--rays, has been
investigated by various authors (Hynes et al. 2000, McClintock et al. 
2001b, Markoff et al. 2001\nocite{Hynes00,McClintock01b,Markoff01}).
The spectrum reported by McClintock et al. (2001b) was obtained from source
observations centered on 2000 April 18 (April 18 in radio, NIR, optical
and X--rays up to 200 keV; April 16--19 in EUV), while that reported by
Hynes et al. (2000) is averaged over observations that took place on
2000 April 2--3 (radio), April 4, 18 (NIR), and April 8 (optical, EUV,
X--rays up to 200 keV). The \sax\ 0.1--200 keV spectrum of the source
performed on 2000 April 14--15 was also reported (F01).
From these spectral  investigations some relevant properties of
the emission processes at the source site can be derived. In the
radio--submillimiter--near-infrared band, the spectrum shows a broken
power--law shape with energy index $\alpha$ 
of $\sim +0.5$ below an inferred cut--off frequency of $\sim 10^{13}$~Hz and 
$\sim -0.6$ above it \cite{Fender01a}. Similar shapes are typical of
hard/low spectral states of BHCs and are believed to be due to synchrotron
emission from self-absorbed jets \cite{Fender01b}. Another relevant
spectral property is the presence of a thermal component consistent with
a blackbody at $\sim40$~eV. This component is apparent in the
first observation of \source\ with \sax\ (F01) and it is also detected in the 
multiwavelength spectrum of the source derived by McClintock et al. (2001b)
\nocite{McClintock01b}. \sax\ has also
allowed  to study the source spectrum up to 200~keV (F01). In the \sax\
observation, the high energy tail
is well fit with a thermal Comptonization in a plasma with a temperature
of $\sim 60$~keV and an optical depth of the order of unity. A weak Compton
reflection  ($\Omega/2\pi = 0.16^{+0.11}_{-0.06}$) along with a low metallicity
($Z/Z_\odot = 0.10^{+0.11}_{-0.06}$) of the reflecting material appears the best 
description of the broad \sax\ spectrum.  Miller et
al. (2002)\nocite{Miller02}, using  0.4--5~keV data obtained with
a {\it Chandra X--ray Observatory} (\chandra) observation performed on
2000 April 18 and 2.8--100 keV data obtained summing together 18 short
\xte\ observations performed between 2000 April 13 and May 13, report a
reflection ($0.01^{+0.06}_{-0.01}$) consistent with zero. This result,
obtained with standard metal abundances ($Z/Z_\odot = 1 $),
is also found  with \sax\ (F01), when the metal abundances are constrained to be
the solar ones.
Markoff et al. (2001) find the multiwavelength spectrum from
radio to X--rays reported by Hynes et al. (2000)\nocite{Hynes00}  also
consistent with nonthermal synchrotron emission from a jet-like matter 
outflow. 

The erratic time variability of the source has also been investigated. As in
several other BHCs in their low/hard state, a low frequency QPO feature
was discovered (Revnivtsev et al.\ 2000, Yamaoka et al.\ 2000).  The centroid frequency is
observed to continuously drift from $\sim 0.07$~Hz to $\sim 0.17$~Hz from 
2000 April 10 to June 11 \cite{Wood00}. It is interesting that
a similar QPO feature as well as a non--Poissonian time variability is found 
in the optical/UV band: the power spectral density estimate (PSD) has a power--law shape
($\propto f^{-\beta}$ with index $\beta$ of $\sim 1$ above 0.02~Hz 
\cite{Haswell00b}. Simultaneous high-time resolution X--ray and optical observations
of the source show a strong correlation between these emissions \cite{Kanbach01}:
the optical emission suddenly (within 30 ms) rises following an increase in the
X--rays, but with a dip in the optical/X--ray cross--correlation at 2--5 s 
before zero lag, as though the optical emission knows something about the X--ray
emission that comes later. Kanbach et al. (2001)\nocite{Kanbach01} interpret these
features assuming a cyclo-synchrotron radiation emission mechanism \cite{Merloni00}
and the photoshere of a magnetically dominated outflow from the black hole  
as the emission region.
Another peculiar fact is that the 0.01--50~Hz fractional variation of the source flux 
is found to be highest in the soft X--ray range (62\% rms in the 0.1--2 keV energy range
versus 42\% rms in the 2--10 keV range)  (F01).

After the first Target of Opportunity (TOO) (F01), we performed with \sax\ three
additional  long observations of  \source\ (see Fig.~\ref{f:outburst}), 
sampling the source outburst and its quiescence status.
In this paper we report results of these observations and compare
them with those, mentioned above, obtained in the first TOO.

\section {Observations and data analysis}
\label{s:obs}

All the observations were performed with the \sax\ Narrow Field Instruments (NFIs).
The NFIs embody a Low Energy Concentrator Spectrometer (LECS, 0.1--10 keV,
Parmar et al.\ 1997\nocite{Parmar97}), 2  Medium Energy Concentrators
Spectrometers (MECS, 1.3--10 keV, Boella et al.\ 1997), a High Pressure Gas
Scintillator Proportional Counter (HPGSPC, 3--100 keV, Manzo et al.\
1997\nocite{Manzo97}), and a Phoswich Detection System (PDS, 15--300 keV,
Frontera et al.\ 1997\nocite{Frontera97}). The LECS and MECS have imaging
capabilities, while the HPGSPC and PDS are collimated detectors with a
Field of View (FOV) of $1^{\circ}$ and $1.3^{\circ}$, respectively.
The PDS instrument makes use of rocking collimators for background monitoring.

The first three observations (see log in Table~\ref{t:log})
were performed in April, May and June 2000, at time intervals of about one
month. During the second observation, the LECS was switched off. 
From 2000 June 30 to October 10, the source was outside the visibility window
for \sax, thus the last \sax\ observation was performed in December 2000, long after
the ASM outburst was over.

Useful data were selected from time intervals in which the satellite
was outside the South Atlantic Geomagnetic
Anomaly, the elevation angle was above the Earth limb by $\geq 5\degr$, the
Earth was dark (only for LECS), and the high voltage supplies were
stabilized.
The LECS and MECS source spectra  were extracted from a region with a radius of $8'$
around the centroid of the source image. As background spectra we used
standard files obtained from the observation of blank fields. For each
observation the spectra of the two MECS were normalized to the same gain and co-added.
We used background spectra accumulated from dark Earth
data for the HPGSPC (note that the instrument's collimator was kept on the
source for the entire observation). The background level of the PDS was
estimated by swapping its collimators off source every 96 s. The energy bands
used for spectral fitting were limited to those where the response functions
were best known, i.e, 0.12--4 keV, 1.7--10 keV, 7--29 keV, and 15--200 keV,
for the LECS, MECS, HPGSPC and PDS, respectively. 
The count rate spectra  were analyzed using the {\sc xspec} software
package \cite{Arnaud96}. A systematic error of 1\% was added in quadrature
to the statistical uncertainties of the spectral data, on the basis of the
calibration results obtained with the Crab Nebula, that was observed
on 10 April 2000, 4 days before the first observation of \source. We
allowed for free normalization of the instruments in
multi-instrument fits with respect to MECS. For clarity of display,
the unfolded spectra from multi-instrument fits were renormalized to the
MECS level.

The quoted errors for the spectral parameters correspond to 90\% confidence
for one parameter ($\Delta \chi^2 = 2.71$). The elemental abundances are
those by Anders \& Ebihara (1982), and the opacities are from
Morrison \& McCammon (1983). For comparison with the first observation
results (F01), we assume the same distance (1.5 kpc) and disk inclination
($i=70\degr$)  assumed in the analysis of the first observation. The parameter values
within square parentheses in the Tables are kept fixed in the fits.

\section{Results}
\label{s:results}

The source was detected in the first three observations, but it was
no more visible in the last one. Separately we discuss the first three
observations (\S~\ref{s:spectra} and \S~\ref{s:time}) and the last one
(\S~\ref{s:too4}).

\subsection{Spectral behavior with time}
\label{s:spectra}

Since the energy spectrum proved to be stable throughout each single observation,
we accumulated spectra corresponding to each of them.
As done in the case of the first observation (see F01), also for the count spectra of the
observations 2 and 3,  a large number of spectral models were tested: 
power--law ({\sc pl}), cutoff power--law ({\sc cutoffpl}), which is equivalent to
the {\sc pexriv} model \cite{Magdziarz95} in {\sc xspec} with no Compton reflection,  
Comptonization model 
({\sc comptt}) of Titarchuk (1994)\nocite{Titarchuk94}, thermal-Compton model ({\sc compps} 
v3.4\footnote{{\sc compps} is available on the internet at 
ftp://ftp.astro.su.se/pub/juri/XSPEC/COMPPS.}) of Poutanen \& 
Svensson (1996)\nocite{Poutanen96}, all either singly taken or with the addition a 
blackbody ({\sc bb}),
all photoelectrically absorbed ({\sc wabs} model in {\sc xspec}). Either spherical (sph) or 
slab (sl) geometries of the Comptonizing electron cloud as well as  solar or free  metal 
abundances $Z$ of the cold medium were tested. 
In the fits with the {\sc compps} model, which allows for the presence of Compton 
reflection \cite{Magdziarz95} from a cold medium, variable element abundances with
respect to the solar ones, could be assumed. For consistency,  
also a narrow iron emission line with Gaussian profile (centroid energy $E_l = 
6.4$~keV and width with $\sigma = 0.01$~keV) was added in the fits. Similarly to 
what found in the TOO1 (F01), also for TOO3 (for TOO2 this
test could not be done given the absence of LECS) the fit results with no addition
of a direct  {\sc bb} are worse in the case of {\sc cutoffpl} and give the same 
goodness in the case of  the {\sc comptt} and {\sc compps} models. But, in the latter case, 
the best-fit $N_{\rm H}$ is found lower or, like in the case of {\sc cutoffpl}, implausibly 
low (see discussion in F01). 
The $\chi^2/{\rm dof}$ obtained for {\sc cutoffpl}, {\sc comptt} and {\sc compps}
with the addition of {\sc bb} are reported in Table~\ref{t:chi2}. 

It is noteworthy  that the {\sc compps} plus 
{\sc bb} model, with $Z/Z_\odot$ free to vary, systemically  provides a
significantly  lower  $\chi^2$  than that in which $Z/Z_\odot$ 
is frozen to 1: using the F--test (e.g., Bevington 1969\nocite{Bevington69}), the 
probability of chance improvement in the case of 
slab geometry is $1.4 \times 10^{-4}$ for TOO1, $2.5 \times 10^{-4}$ for TOO2 and $4.7 
\times 10^{-8}$ for TOO3. In  the case of a spherical geometry, 
the probability values are higher but still low, specially for observation 3 ($3.7 \times 
10^{-5}$).
Thus we consider this two--component model with non solar metal abundance 
$Z$ of the cold medium (likely an accretion disk), the most 
suitable model capable of describing the spectra of \source.  The blackbody 
seed photons are assumed to be homogeneously distributed in the plasma.
Either spherical or slab geometries of the Comptonizing  electron cloud fit the data 
(see Table~\ref{t:chi2}).  In Table~\ref{t:results_1} we show the best fit results 
in the case a slab geometry (F01). The Thomson optical depth $\tau$ corresponds
to the half-thickness of the slab. The metal abundance $Z$ is about a factor 10 lower
than the solar. In observation 1 the best fit absorption column density $N_{\rm H}$ is
consistent with the Galactic column  of H{\sc i} measured in radio in the
direction to the source ($\nh\simeq (1.28$--$1.44)\times 10^{20}$
cm$^{-2}$, Dickey \& Lockman 1990), but it is lower by a factor $\sim 2$
in the last observation. We observe a systematically 
significant Compton reflection, which is not required (2$\sigma$ upper limits of 
0.15, 0.09 and 0.08 for TOO1, TOO2 and TOO3, respectively), when solar metal abundances 
are assumed.
The Compton reflection is consistent with zero even if only the Fe abundance is left
free to vary in the fit (in this case $(Z/Z_{\odot})_{\rm Fe}$ converges to 0.1).
Using the {\sc pexriv} model, in place of {\sc compps}, to compare our 
results with those by Miller et al. (2002)\nocite{Miller02}, 
a Compton reflection is requested by the data ($\Omega/2\pi \approx 0.2$ independently
of the observation, with an uncertainty of about 0.1 at 90\% confidence level) even 
in the case of solar metal abundances ($\chi^2/{\rm dof} =$ 206.8/175, 108.3/107 and
189.1/167 for TOO1, TOO2 and TOO3, respectively) . However we find  that, also 
with this model,
a better fit is found, specially for TOO3, leaving  $Z/Z_{\odot}$ free to vary 
(probability of chance improvement of $3.7 \times 10^{-2}$ for TOO1,  $5.2 \times 10^{-2}$
for TOO2 and $6.0 \times 10^{-6}$ for TOO3). A Compton reflection component 
is still needed in this case with a mean metal abundance of $0.27\pm 0.09$.

\begin{figure}[!t]
\epsscale{0.8}
\plotone{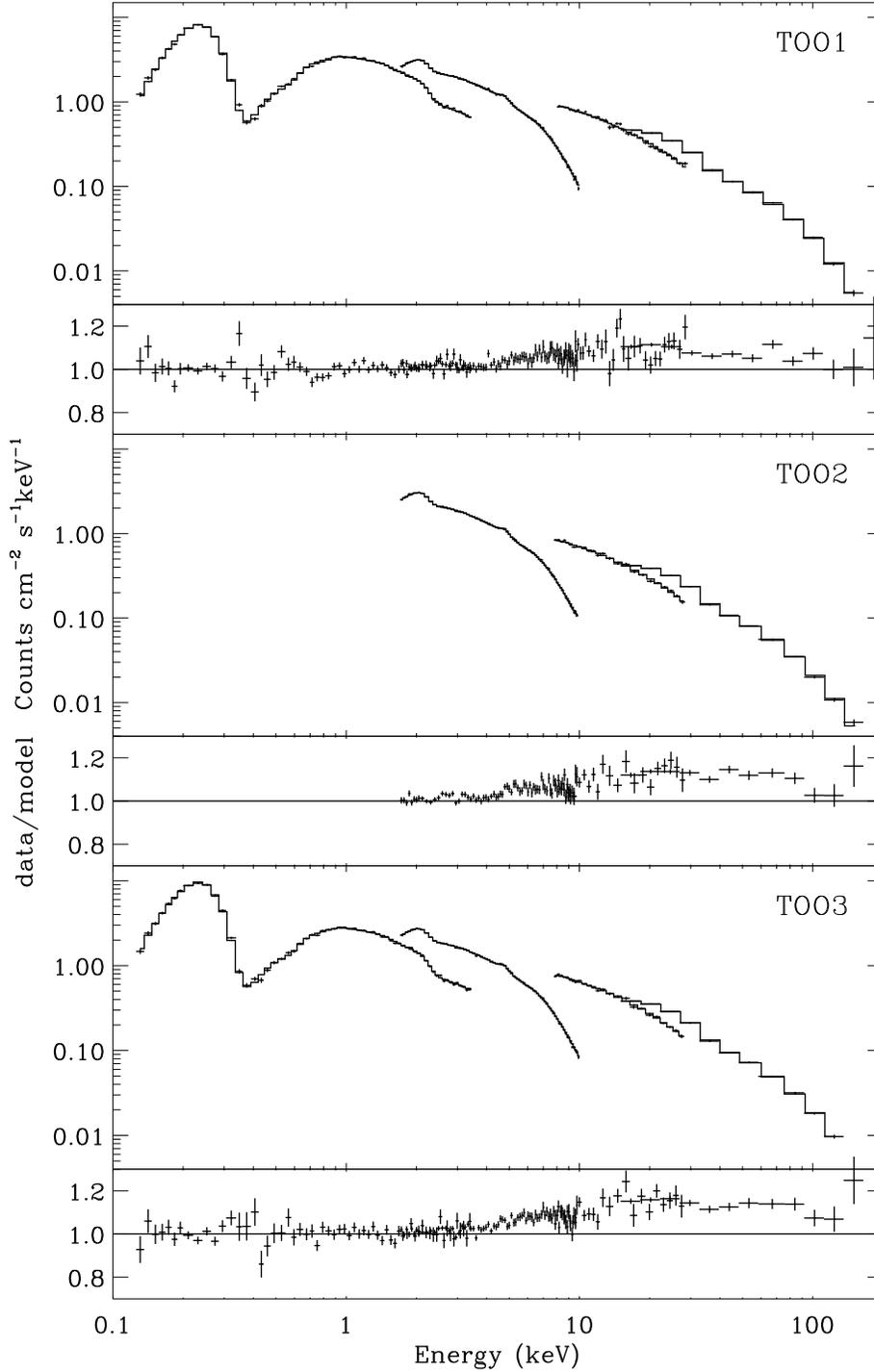}
\vspace{0.5cm}
\caption{{\it Top side of each panel}: The \sax\/ count rate spectra with 
superposed ({\it solid line}) the best fit model ({\sc bb} plus {\sc compps}) 
for a slab geometry of the Comptonizing electron cloud and with metallicity of the reflecting 
medium free to vary (see  Table~\ref{t:results_1}). 
{\it Bottom side of each panel}: residuals to the  model after removing the 
Compton reflection component from the model ($\Omega/2\pi = 0$ is assumed). The 
signature of a reflection component is apparent.} 
\label{f:zero_refl} 
\end{figure}

The count rate spectra  and the best fit model ({\sc compps} plus {\sc bb} for
a slab geometry and with reflection, $Z/Z_\odot$ free to vary) are shown 
in Fig.~\ref{f:zero_refl}. In the same figure the residuals to the best fit model
are shown when the Compton reflection component is removed from the model. 
As can be seen, the effect of this removal is apparent.

The {\sc compps} parameters do not show statistically relevant changes with time. 
The best-fit temperature of the Comptonizing electrons increases from $\sim 60$~keV 
to $\sim 90$~keV, but, within the 90\% errors, these values are consistent with 
each other. Also the Compton reflection, metal abundance and 
optical depth of the electron cloud show statistically negligible changes.
However the temperature of the direct {\sc bb} emission increases and its luminosity 
decreases from observation 1 to 3, with the radius of the {\sc bb} emission
region decreasing from $\sim 35$ to $\sim 8 R_g$\footnote{$R_g \equiv G M_X/c^2$, 
assuming $M_X = 10 M_\odot$}.
In observation 2, given the absence of the 
LECS data, the addition of a {\sc bb} does not influence the fits. 
No evidence of an iron fluorescence K line is found in the observations from 1 to 3, 
with a $2\sigma$ upper limit of the equivalent width ($EW$) of $\sim 30$~eV. This
result is consistent with the low Compton reflection level.

\begin{figure}[!t]
\epsscale{0.8}
\plotone{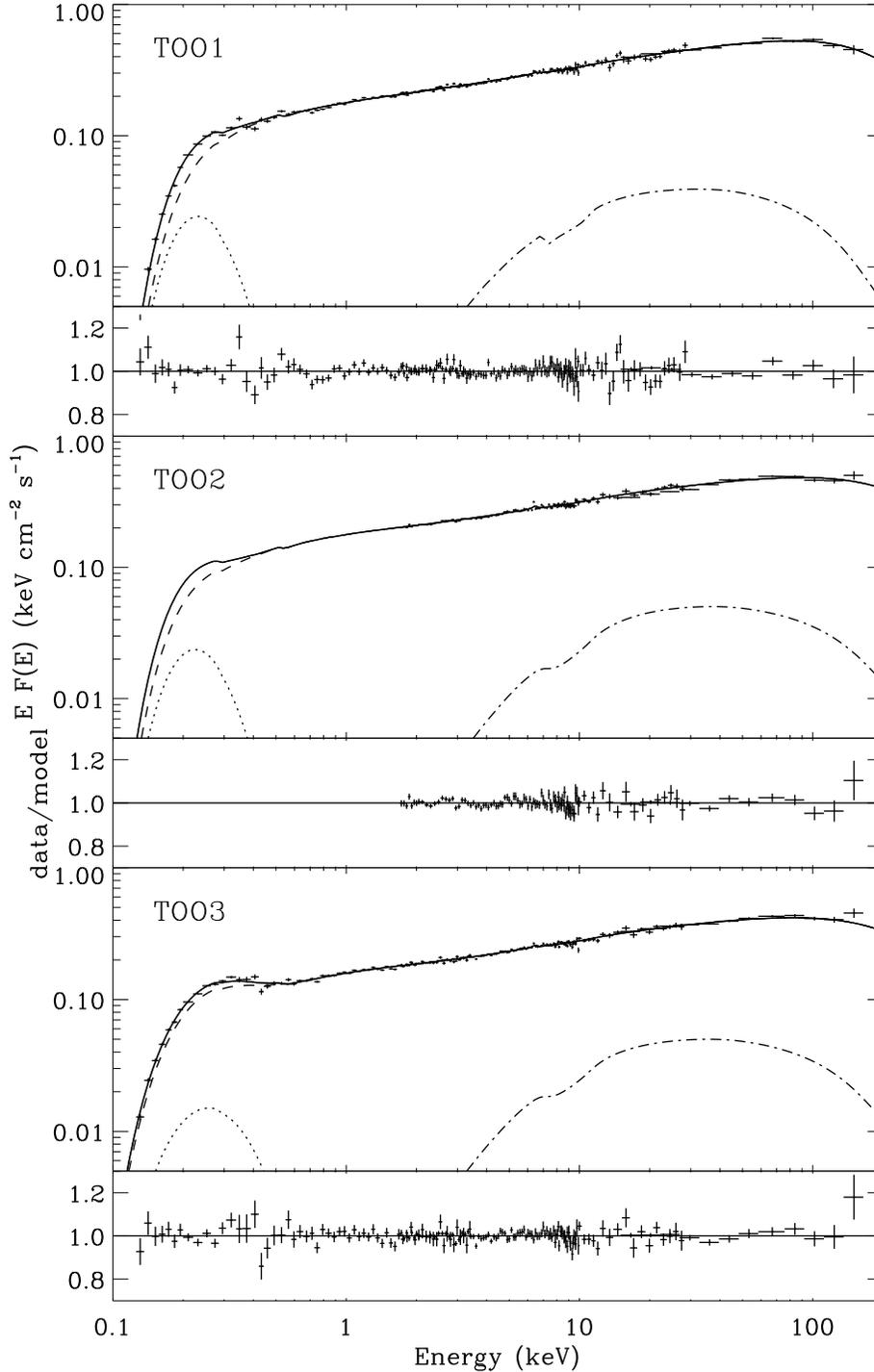}
\vspace{0.5cm}
\caption{The \sax\/ spectra ({\it crosses}) fitted by ({\it solid curve}) a {\sc bb} plus
{\sc compps} model assuming a slab geometry of the Comptonizing electron cloud and the 
metallicity of the reflecting medium free to vary (see Table~\ref{t:results_1}). 
The {\sc compps} best fit model is shown separately by  dashed curves
while the direct {\sc bb} is shown by dotted lines. The dot--dashed curve gives the 
reflection component. The systematic presence of a reflection component is apparent, 
while the {\sc bb} component is  reduced in the third observation. The spectra are 
normalized to the level of the MECS (see text). The bottom side of each panel shows 
the fit residuals.
} 
\label{f:compps} 
\end{figure}

\begin{figure}[!t]
\epsscale{0.8}
\plotone{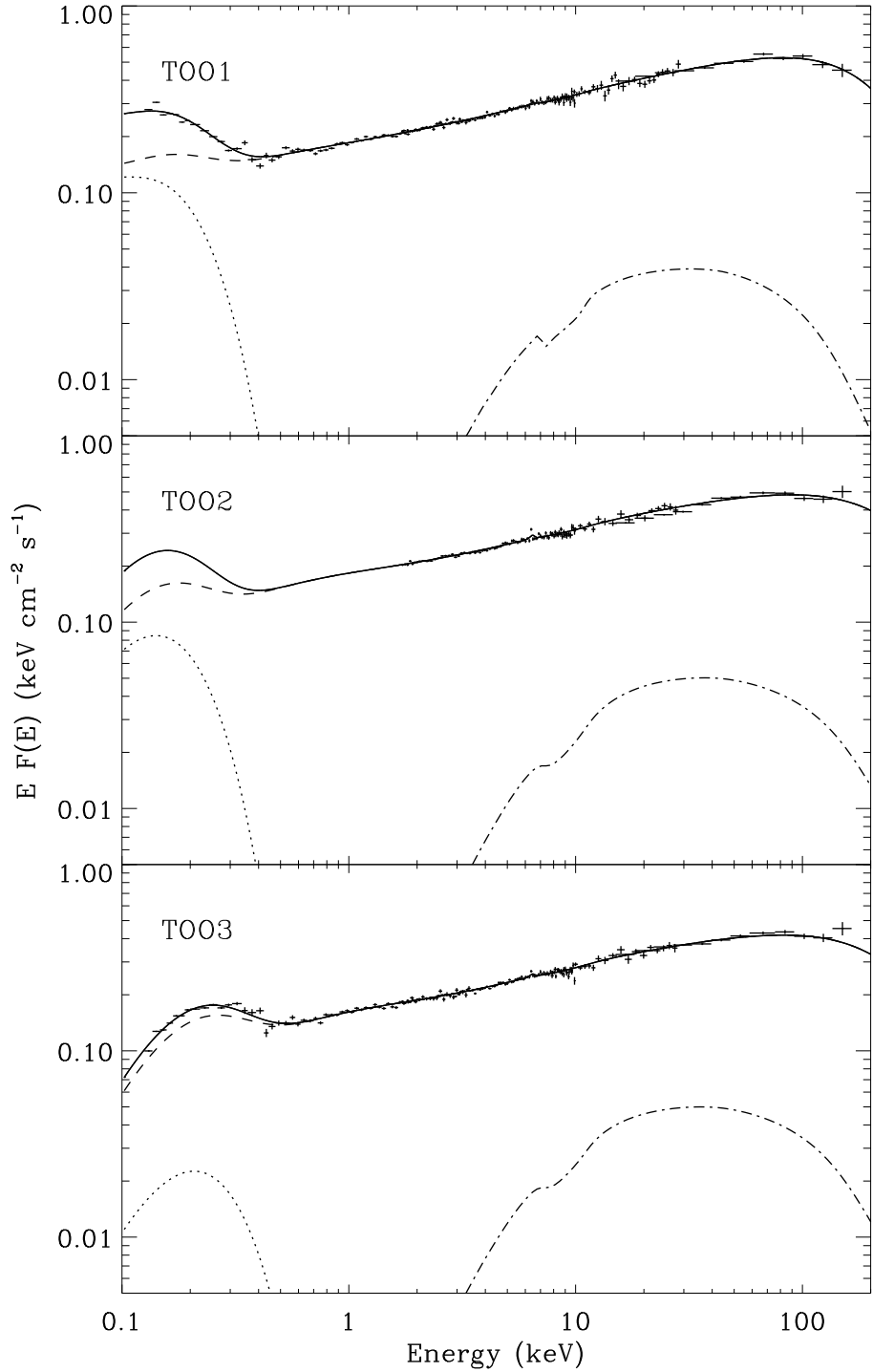}
\vspace{0.8cm}
\caption{The absorption-corrected model spectra (solid curve)
corresponding to the fit shown in Fig.\ \ref{f:compps}. The initial
(i.e., before Compton scattering) spectrum of the seed photons incident on
the plasma corresponds to  the peak at $\sim$0.15~keV. (see also Table~\ref{t:results_1}).
} 
\label{f:EFE}
\end{figure}

The best fit results with the photoelectrically absorbed and unabsorbed {\sc compps} 
plus {\sc bb} with free $Z$ and slab geometry are shown in  Fig.~\ref{f:compps}
and Fig.~\ref{f:EFE}, respectively. The small residuals to the best fit model 
are apparent.

As in the case of TOO1 (see F01), we have also considered for TOO2 and TOO3 a 
multi-component model consisting of the {\sc compps} model in which the spectrum of 
seed photons peaks below the observed range ($kT_{\rm seed}$ fixed at 10 eV), plus an 
additional {\sc bb} component plus a narrow Gaussian profile at 6.4~keV. 
Such a low energy of seed photons is expected, e.g., when Comptonization of thermal 
synchrotron
emission dominates (e.g.\ Wardzi\'nski \& Zdziarski 2000). This model, in both 
geometries (sphere or slab), yields a good description of the data (see Table~\ref{t:chi2}).
The best fit parameters for the three TOOs are reported in Table~\ref{t:results_2}.
As can be seen, also in this case the electron temperature appears to increase
and the luminosity of the {\sc bb} component to decrease from TOO1 to TOO3.

Leaving the metallicity free to vary, as done for observation 1 (see F01), also for
observation 3 (for the second observation it could not be done due to the
absence of the LECS data) we have tested the effect of replacing a {\sc bb} with a disk 
blackbody, 
in this case assuming the seed photons for Comptonization to be also those of the disk
blackbody. This model for the third observation yields plasma parameters very similar to
the blackbody case above and a similar fit quality ($\chi^2/{\rm dof} = 169/165$
for a slab and $= 175/165$ for a sphere). The inner disk temperature and radius
are still consistent with the values found in TOO1: $kT_{\rm in}=41^{+3}_{-4}$~eV, 
$R_{in}\ga 500$~km. The lower limit of $R_{in}$ corresponds to $\sim 34 R_g$.

The mean flux levels of the source  in the three observations are reported in 
Table~\ref{t:results_1} for three different
energy ranges. Correspondingly, the bolometric (0.01--$10^3$ keV) luminosity, corrected 
for absorption, decreases with time from  
$1.1 \times 10^{36} \,$\ergs\  to $9.9 \times 10^{35} \,$\ergs.
 
\subsection{Temporal behavior}
\label{s:time}

\begin{figure}[!t]
\epsscale{0.8}
\plotone{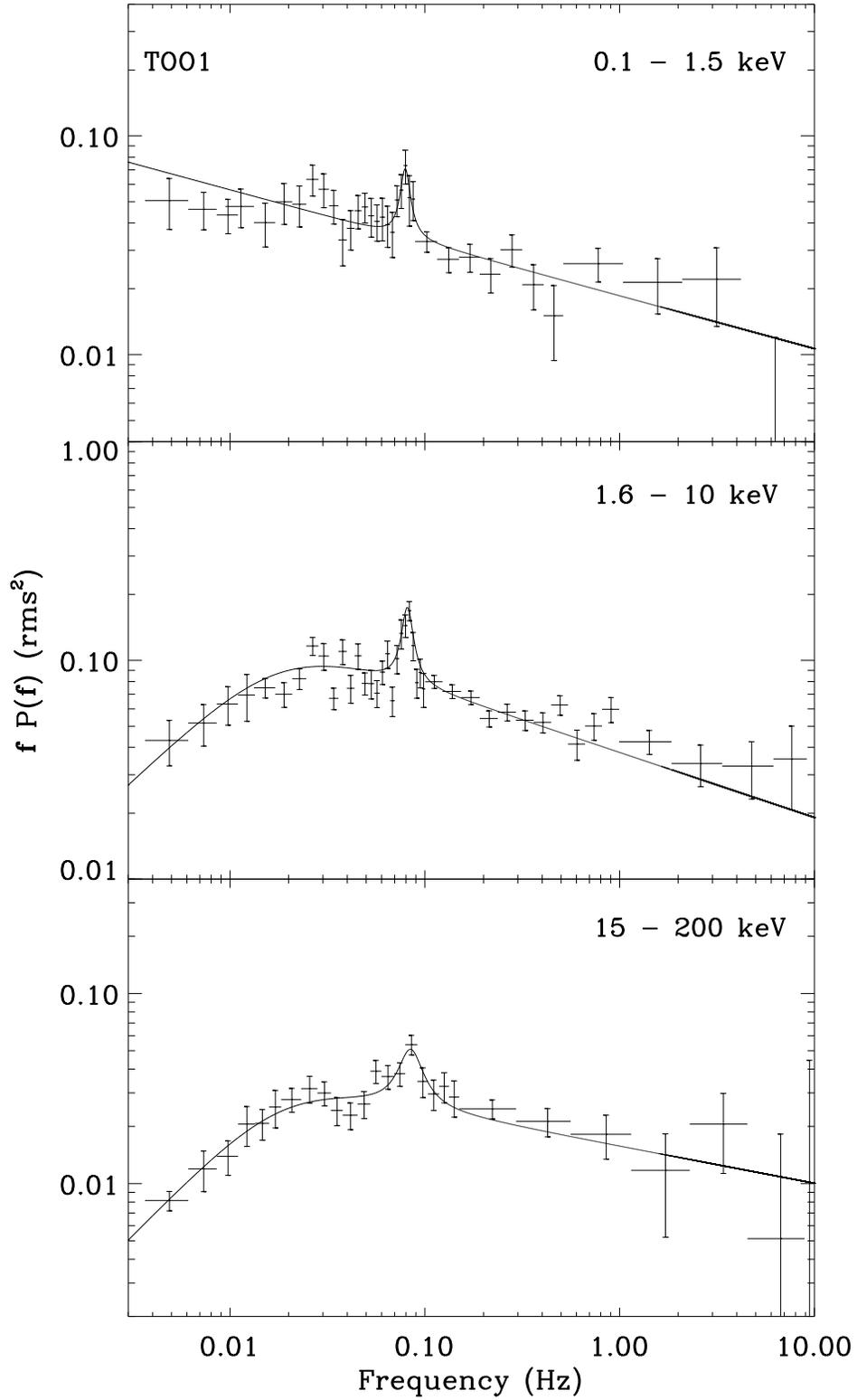}
\vspace{0.8cm}
\caption{Fourier frequency times Power Spectral Density  of the source flux 
variations during TOO1 versus frequency in  the  0.1--1.5 keV, 
1.6--10 keV and 15--300 keV energy bands. The Poissonian variance 
is subtracted. The QPO is apparent.}
\label{f:psd_t1}
\end{figure}

\begin{figure}[!t]
\epsscale{1.0}
\plotone{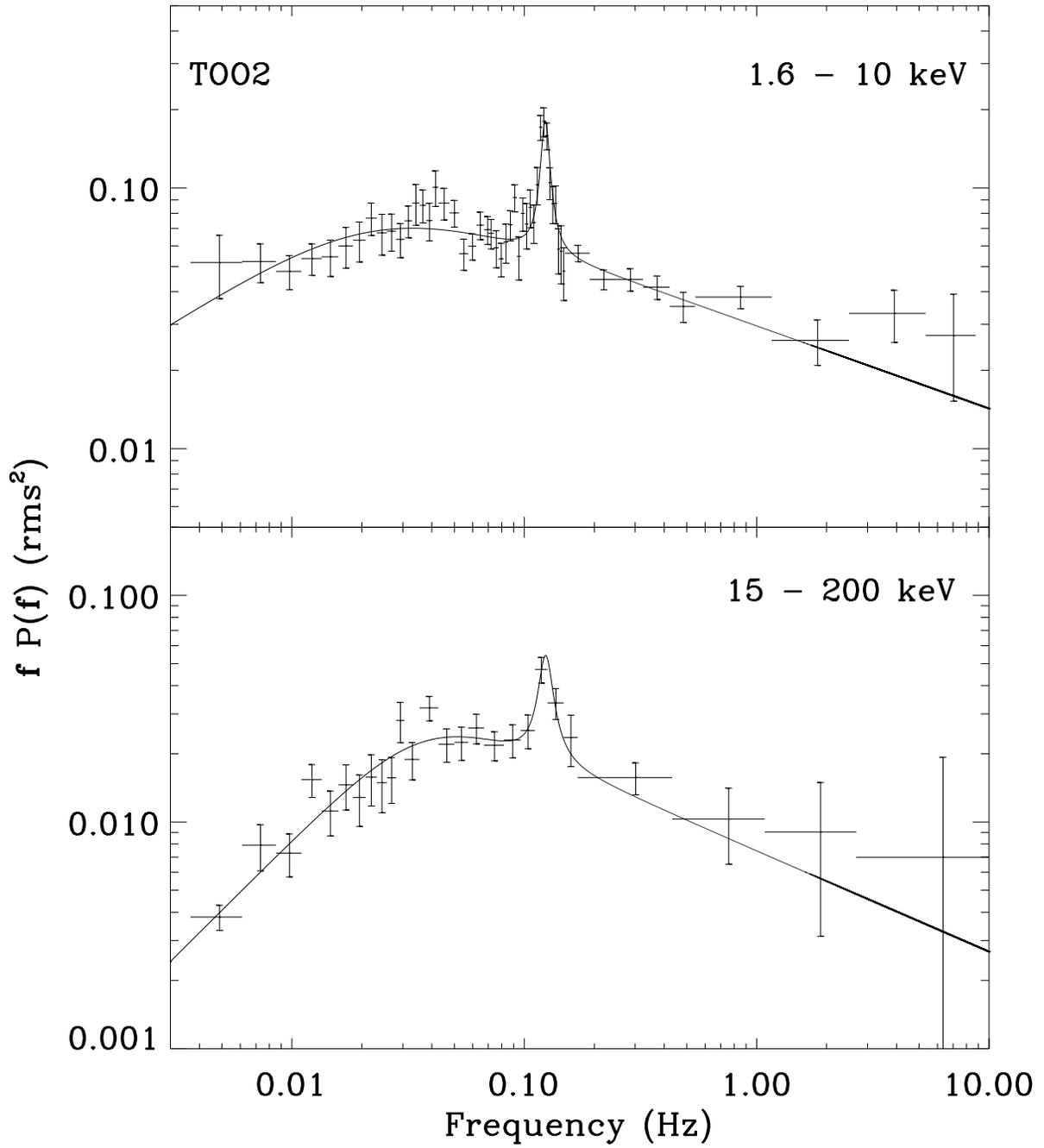}
\vspace{-2.5cm}
\caption{Fourier frequency times Power Spectral Density  of the source flux 
variations during TOO2 versus frequency in  the  1.6--10 keV and
15--200 keV energy bands. 
The PSD in the 0.1--2 keV is lacking because LECS was switched off. The 
Poissonian variance is subtracted.}
\label{f:psd_t2}
\end{figure}

\begin{figure}[!t]
\epsscale{0.8}
\plotone{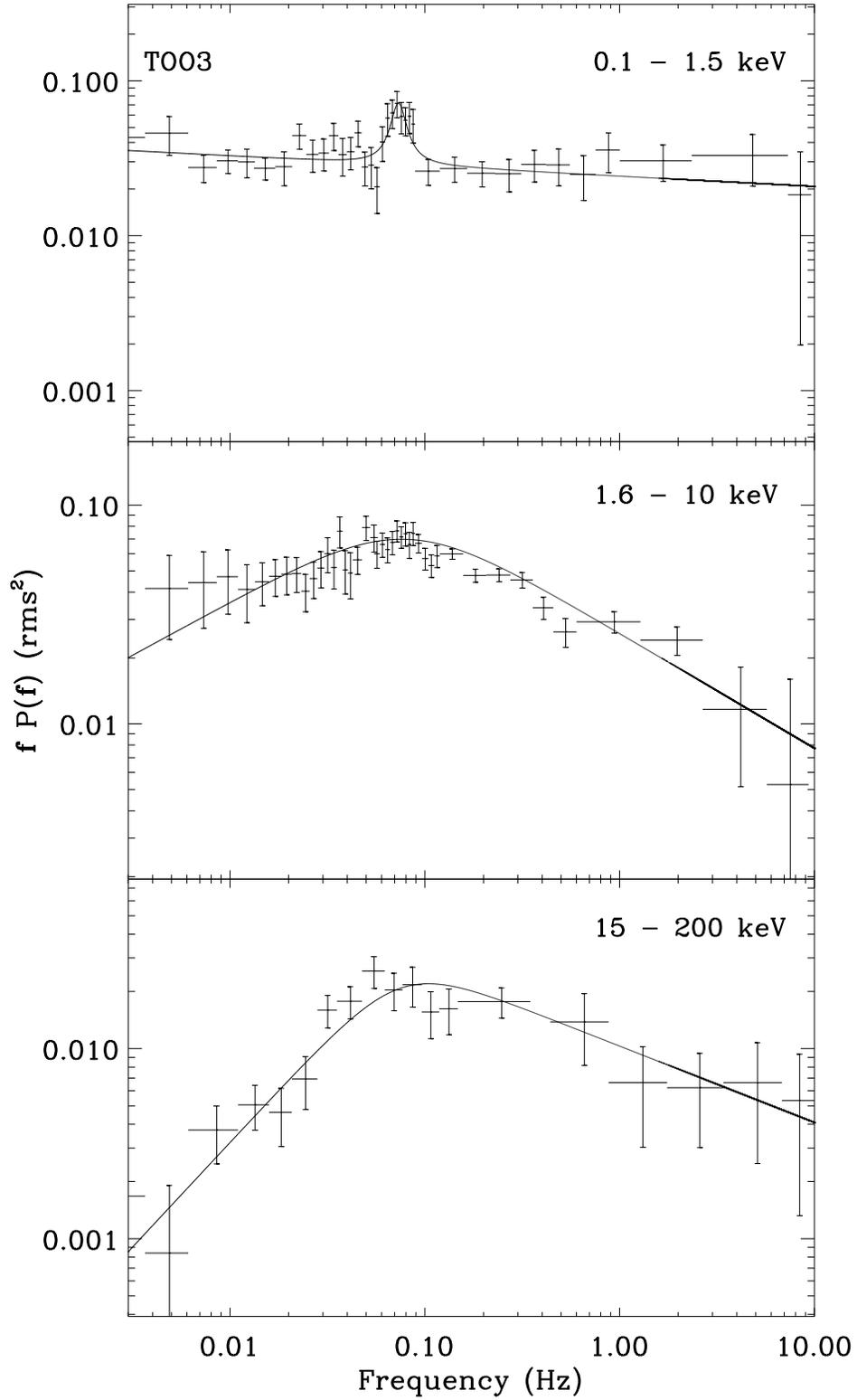}
\vspace{0.8cm}
\caption{Fourier frequency times Power Spectral Density  of the source flux 
variations during TOO3 versus frequency in  the  0.1--1.5 keV,
1.6--10 keV  and 15--200 keV energy bands. The Poissonian variance 
is subtracted.}
\label{f:psd_t3}
\end{figure}

\begin{figure}[!t]
\epsscale{1.0}
\plotone{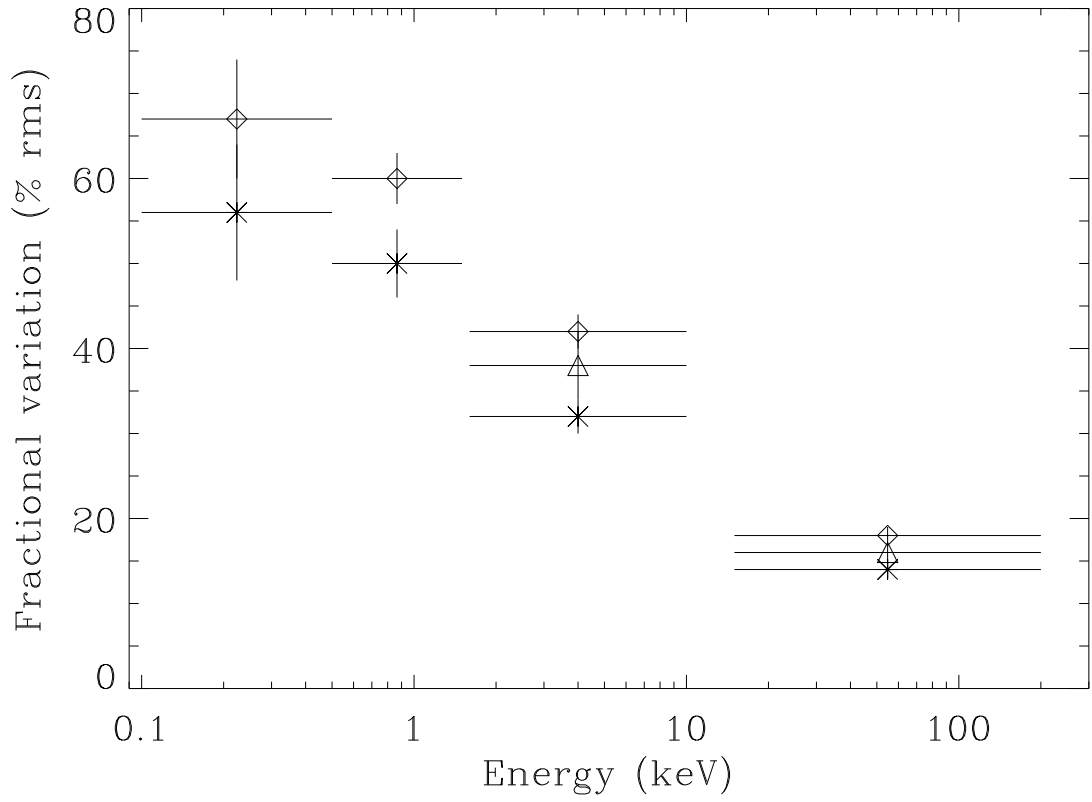}
\vspace{0.8cm}
\caption{Fractional variation of the source flux with energy in the three TOOs.
{\it Diamond:} TOO1; {\it triangle:} TOO2; {\it asterisk:} TOO3.}
\label{f:rms}
\end{figure}

A temporal analysis of the data with Fourier techniques was performed for each of
the first three TOOs, where the source was detected. We estimated the PSD in four 
energy intervals (0.1--0.5, 0.1--1.5, 1.6--10, 15--200 keV) 
for each of the TOOs in the 0.0035--50~Hz frequency range of the time variations.
For all data we adopted a time binning of 0.05~s and data stretches $\sim 410$~s long. 
The number of stetches ranged from $\sim 40$ to $\sim 120$, depending
on the specific observation and instrument considered. PSDs had been produced 
from each of the stretches and averaged over the number of data stretches available 
for the given TOO and energy band. When necessary, the average PSDs were also
rebinned in order to increase the source signal. The white-noise level due to 
Poissonian statistics was subtracted. The PSDs were normalized such that its integral
gives the squared rms fractional variability of the source. 
The derived PSDs multiplied by its Fourier 
frequency $f$ are shown in Figs.~\ref{f:psd_t1}, \ref{f:psd_t2} and \ref{f:psd_t3} as a 
function of $f$. 
The QPO feature first reported by Revnivtsev et al. (2000)\nocite{Revnivtsev00} is  
apparent in most of the estimated PSDs. Each average spectrum $P(f)$ was thus  fit 
with a Lorentzian 
plus a smoothly broken power--law function (e.g., Lazzati 2002 \nocite{Lazzati02}) 
plus a constant $K$:
\begin{eqnarray}
P(f) & = & \frac{R_{QPO} \Delta_{QPO}}{2\pi} \frac{1}{\left(f-f_{QPO}\right)^2+ \left(\frac{\Delta_{QPO}}{2}\right)^2} 
           + \nonumber \\
     &   & \frac{2 F_0}{\left[\left(\frac{f}{f_0}\right)^{2\alpha_1} + \left(\frac{f}{f_0}\right)^{2\alpha_2}\right]^{1/2}} + K
\end{eqnarray}
where $\Delta_{QPO}$ is the FWHM of the fundamental frequency $f_{QPO}$ of the QPO, $R_{QPO}$ 
is the fractional variation of the QPO, $f_0$ is the break frequency, $\alpha_1$ is
the PSD slope below $f_0$, $\alpha_2$ is the slope at higher frequencies, $F_0$ is the
value of $P(f)$ at $f_0$, and $K$ is the contribution of white noise above the Poissonian
variance.  $K$ was found to be consistent with zero. All fits were acceptable.
However, in the 0.1--0.5 keV and 0.1--1.5 keV energy bands, where there is no evidence of a
break (see Figs.~\ref{f:psd_t1} and \ref{f:psd_t3}), $f_0$ could not be 
constrained and the fit was done with a simple power--law of slope $\alpha_2$.
The best fit parameters  and derived quantities (fractional variation $R_t$ of the 
total source flux, $R_{QPO}$  and their ratio)   are summarized in 
Table~\ref{t:time}.
Some interesting features can be derived from these results. The power law index above 
the break
is almost stable ($\sim 1.3$), except in the 0.5--1 keV range, where changes from $\sim 1$ 
in observation 1 to $\sim 1.7$ in the third observation. A flattening of the PSDs 
at low frequencies is observed in all energy bands except below 1.6 keV, as it is 
also apparent in the $f P(f)$ spectra, which must show a maximum corresponding
to a break.  
The behavior of the fractional variation of the source flux with energy is 
shown in Fig.~\ref{f:rms}.
$R_t$ is highest in the 0.1--0.5 keV energy range, where the {\sc bb} component shows 
its maximum contribution. However it  
decreases from the first (67\% rms)  to the third observation, where it gets a 
value of 56\% rms. At higher energies, $R_t$ is lower and similarly decreases with time
like its time behavior in the 0.1--0.5 keV band. A peculiar behavior is shown by
the centroid frequency of the QPO: it rises from $\sim 0.08$~Hz to  $\sim 0.12$~Hz 
from TOO1 to TOO2, in agreement with the results 
reported by Wood et al. (2000)\nocite{Wood00}, but returns back
to $\sim 0.075$~Hz in the June 26 observation. In addition (see
Figs.~\ref{f:psd_t1}, \ref{f:psd_t2} and \ref{f:psd_t3}), while in the 0.1--1.5 keV 
energy range the QPO feature is apparent in both TOO1 and TOO3, in the higher
energy bands it is not visible in the third observation. Similarly to the fractional
variation of the continuum $R_t$, the fractional variation under the QPO peak $R_{QPO}$ 
decreases with energy. However  its relative contribution with respect to the
the total power is approximately stable with energy and time (see Table~\ref{t:time}).

\subsection{Source flux upper limit of the December 2000 observation}
\label{s:too4}

No statistically significant X--ray emission was detected in the long TOO observation
of December 2000 (TOO4). The 2$\sigma$ upper limit of the source flux in the 0.1--2,
2--10 and 15--200 keV energy bands is given by $1.2 \times 10^{-13} \,$\ergcms,
$4.9 \times 10^{-14} \,$\ergcms and $1.9 \times 10^{-11} \,$\ergcms, respectively.
Assuming a distance of 1.5 kpc, the corrisponding 2$\sigma$ upper limit of the
source luminosity is $3.2 \times 10^{31} \,$\ergs, $1.3 \times 10^{31} \,$\ergs,  
$5.1 \times 10^{33} \,$\ergs, respectively.
These upper limits are  in agreement with the luminosity estimates  derived for other
black hole candidates with the {\it Chandra} satellite \cite{Garcia01}.

\section{Discussion}
\label{disc}

The bolometric luminosity undergoes a small decrease ($\sim 10$\%) in three months 
(April--June 2000)  while the spectral shape remains essentially constant.
The spectral shape is  compatible with a direct blackbody  plus a
Comptonization component due to electron scattering in a hot central disk of seed photons. 
Two possible origins for the seed photons are found to be consistent with our data. 
One is that they
come from the outer disk and thus that they are due to {\sc bb} emission, part of
which we directly observe. We find that its temperature  
increases with time from $\sim 36$ to $\sim 52$~eV and its luminosity correspondingly 
decreases (see Table~\ref{t:results_1}). 

The other possibility is that the seed photons  are due to synchrotron of thermal 
electrons located in the hot plasma itself. Indeed, assuming a photon temperature of 
10~eV, corresponding to the turnover frequency above which the plasma becomes
optically thin to the synchrotron radiation, the fits are equally good 
(see Tables~\ref{t:chi2} and \ref{t:results_2}). As discussed by F01, in 
this case the soft excess, visible in the three TOOs, can represent either
the peak of the self-absorbed thermal 
synchrotron emission (e.g.\ Wardzi\'nski \& Zdziarski 2000) or blackbody emission not 
related to the dominant source of the seed photons. The low source luminosity ($L_{\rm
X}/L_{\rm E}\la 10^{-3}$) is compatible with  the synchrotron scenario 
\cite{Wardzinski00}. 
The temporal properties of the source flux  can help to understand the nature of the 
seed photons and that of the soft photons (0.1--0.5 keV) directly observed. 
The derived PSD spectra (see Figs.~\ref{f:psd_t1}, \ref{f:psd_t2} and 
\ref{f:psd_t3} and Table~\ref{t:time}) and the fractional variation
(see Fig.~\ref{f:rms} and Table~\ref{t:time}) of the source time 
variability   show  the presence of a significant level of non-poissonian noise. This 
noise is particularly strong at the lowest energies, where the fractional variation
achieves its maximum value (67\% in TOO1 and 56\% in TOO3). A non-poissonian 
variance is not expected in the case of a {\sc bb} radiation, as also 
confirmed in the case of high/soft states of BHCs (see, e.g., GX339$-$4, Belloni et al. 
1999\nocite{Belloni99}), where the thermal component dominates the spectrum and
there is little variability observed.  Thus a mere {\sc bb} is  inconsistent 
with the source variability properties, while the addition of a non-thermal process 
(likely synchrotron) appears more likely. The optical flickering 
(about 20\% of the total optical flux), the 0.08 Hz QPOs also observed in the 
optical band and the very flat $E F(E)$ spectrum from the NIR to
the UV \cite{Haswell00b,Chaty01,Kanbach01} confirm this picture, without
excluding the contribution to the total soft X-ray flux of a thermal component
\cite{McClintock01b}.
%from the presence of Balmer jump in absorption at $\approx 8 \times ^14$~Hz. 
The inverted power-law spectrum of the radio emission  
definetely points to  a synchrotron radiation component in this band \cite{Fender01a}.

Asssuming a synchrotron origin for the soft X--ray excess, in F01 we 
constrained the source magnetic field intensity to be $3\times 10^6 \la B 
\la 10^8$~G. We confirm that the data of the second and third TOOs give the
same constraint for $B$. 

The best-fit values  of the Comptonizing electron temperature increase with time 
from $\sim 60$ to $\sim 100$~keV; however, due to their  statistical uncertainties, 
we cannot conclude that the this temperature actually change with time.
The optical thickness of the electron cloud appears stable, within errors. We 
definitely observe  radiation which is Compton reflected by a medium 
(likely the accretion disk around the BH) with metal abundance lower, by 
about a factor 10, than that 
of the solar environment.  The reflection is not detected if we assume a solar
metallicity, but this assumption  provides a worse description of the data. Our
results found with the {\sc compps} model are not in disagrement with those obtained 
by Miller et al.  (2002), who instead used the {\sc pexriv} model, finding a 
Compton reflection consistent with zero assuming either a solar metallicity 
or $(Z/Z_\odot)_{Fe}=0.1$.
However with {\sc pexriv} we find a Compton reflection component even assuming a
solar metallicity. Likely the higher accuracy of our spectra at high energies is at the 
origin of the difference between our results and those by Miller et al. (2002).
The low metallicity derived by us (mean value of $Z/Z_\odot = 0.13^{+0.06}_{-0.04}$) 
is expected in the case of a halo object, as it is  \source. Haswell 
et al. (2002)\nocite{Haswell02}, analyzing ultraviolet observations of the source,  
find a very high underabundance of carbon with respect to the other metals  and
do not exclude a lower abundance of these metals  with respect to that typical of
a solar--like environment. 
No evidence of an iron line is found down to $\sim 30$~eV. As we discussed in our 
first paper (F01), the weakness of the Compton--reflection components 
could be naturally accounted for by a small solid angle subtended by the cold 
medium as seen from the hot disk, but could also be explained by the sub-solar
abundances derived. 
Esin et al. (2001), in the context of the the Comptonization model propose
that the hot central disk is advection--dominated. As discussed by F01,
if such interpretation is correct, it is not clear how to compare 
the spectral results found for \source\  with those obtained for Cygnus X--1, 
which has a much larger luminosity and almost the same spectral parameters 
\cite{Esin98,Gierlinski97,Frontera01a}.

Markoff et al. (2001)\nocite{Markoff01} propose that not only the
radio--optical but also the X--ray emission could be due to non-thermal jet synchrotron. 
According to
these authors, the radio to optical power output  is attributed to optically thick 
synchrotron emission from the jet at distances from the BH higher than the region where 
the charged particles are shock accelerated 
($z>z_{acc}\approx 100 R_g$, while the X--ray output is due to 
optically thin synchrotron emission at the shock
acceleration region ($z \sim z_{acc}$). On the basis of this model, the
spectrum expected in the hard X--ray band is a  power--law with a cutoff.
The exact shape of the cutoff has not been fully worked out for this model.
They base their conclusion on the \source\ emission model on the spectrum up 
to 100 keV derived from the {\it High Energy X--ray Timing
Experiment} (HEXTE) aboard \xte\ \cite{Hynes00}, which however does not show any cutoff. 
The \sax\ data demonstrate the presence of this feature, but also show that a power--law 
with an exponential cutoff does not describe the data.
In addition a Compton reflection component, even if weak, is apparent in the data.
These facts put strong constraints to the synchrotron model proposed by these authors. 
At this state of knowledge, while a synchrotron emission component appears 
consistent with the low energy data (below 1 keV), an Inverse Compton scattering 
component appears  to be the
best scenario for fully describing the hard X--ray energy band.

\section{Conclusions}

The  column density measured in the first \sax\ observation is consistent with 
the Galactic $N_{\rm H}^{\rm G}$ obtained from 
radio measurements in the direction to the source ($\nh\simeq (1.28$--$1.44)\times 
10^{20}$ cm$^{-2}$, Dickey \& Lockman 1990). However the $N_{\rm H}$
derived from the June  observation (TOO3) is lower by about a factor 2. Likely a local 
absorption contributes to the absorption measured during the April observation (TOO1). 
The centroid frequency  
of the source QPOs is found higher during the second observation, consistently with the
monitoring results  found by Wood et al. (2000)\nocite{Wood00}, but we
find that in the third observation (2000 Jume 26), unexpectedly it again decreases,
going back to  the April 12 value \cite{Wood00}. 

The source spectra  show  a soft excess which is consistent with a {\sc bb} 
emission with temperature which increases with time from $35^{+3}_{-4}$~eV (TOO1)
to $52^{+7}_{-6}$~eV (TOO3), even though the power spectra
flux variations do not favour this interpretation.
The high energy part of the spectra is well fit by a 
Comptonization in a thermal plasma with stable Thomson 
optical depth of $\tau\sim 1$ and electron temperature which does not
significantly change with time (mean value of $T_e = 84^{+13}_{-15}$. 
The values of $kT_e$ and $\tau$ we find are very similar to those typical of more
luminous  black-hole binaries in the hard state (e.g., $L_{\rm X}\sim 0.02 L_{\rm E}$ 
in Cyg X-1), whereas the X-ray luminosity in our measurement is
only $\sim 10^{-3}L_{\rm E}$.  
A Compton reflection component is apparent and stable, even if weak 
(mean value of $\Omega/2 \pi = 0.21^{+0.05}_{-0.04}$). 
This component is only detected when the reflector metallicity is left free to vary
in the fits, but the \sax\ data, in particular those of the third observation,
definitely point to a reflector with a non-solar metallicity 
(mean value of $Z/Z_\odot = 0.13^{+0.06}_{-0.04}$), as expected by an environment 
at high galactic latitude, as it is the case for \source. Our results are 
consistent with the assumption,
discussed by Mirabel et al. (2001)\nocite{Mirabel01}, that \source\ is the relic of an
ancient massive star formed in the Galactic halo. 
No Fe $K \alpha$ line is 
detected ($2\sigma$ upper limit of $\sim 2 \times 10^{-4}$~cm$^{-2}$ s$^{-1}$). 

On the basis of the fit results, a hot central disk appears the best scenario for
the high energy photons. As far as the origin of the soft photons is concerned, the low 
amplitude of reflection and the iron line intensity would point to the assumption that they
are {\sc bb} radiation from the outer cold disk. However, taking into account the temporal 
properties, in particular the non-poissonian variance of the time variations of the soft
X--ray flux, likely a relevant part of them have a different origin, likely synchrotron 
emission internal to the hot disk.

\acknowledgements

We wish to thank Sera Markoff for  useful comments to the paper. 
The \sax\ satellite is a joint program of Italian (ASI) and Dutch (NIVR) 
space agencies. We acknowledge  ASI for supporting this research. FF also 
aknowledge a 
financial support from the Ministry of University and Scientific Research of Italy 
(COFIN 2000 and 2002).
AAZ acknowledges support from the Foundation for Polish Science and KBN grants
5P03D00821, 2P03C00619P1,2 and PBZ-KBN-054/P03/2001.

\clearpage

%
% Table 1
%

\begin{deluxetable}{l c c c c c c c c c}
\tabletypesize{\small}
\tablecolumns{10} 
\tablewidth{0pc}
%\tablenum{1}
\tablecaption{Log of \sax\ TOO observations of XTE J1118+480}
\tablehead{
\colhead{TOO} &\colhead{Epoch} & \multicolumn{4}{c}{Exposure time (ks)} & \colhead{} & \multicolumn{3}{c}
{Mean rate (counts s$^{-1}$)} \\
\# &  & \colhead{LECS} & \colhead{MECS} & \colhead{HP} & \colhead{PDS} & \colhead{} & 
\colhead{0.1--2 keV} & \colhead{2--10 keV} & \colhead{15-200 keV}
}
\startdata
1 & 2000 April 14.49--15.46 & 21   & 30  & 420  & 20  &  &   $5.181 \pm 0.016$ &
   $7.466 \pm 0.017$  &   $ 12.17 \pm 0.05$ \\ 
2 & 2000 May  24.20--25.18  & --   & 40  & 37   & 19  &  &     -- &
   $7.136 \pm 0.013$  &   $10.98 \pm 0.05$   \\
3 & 2000 June 26.46--27.70  & 16   & 48  & 47   & 20  &  &  $4.586 \pm 0.018$  &
   $6.35 \pm 0.01$  &     $9.85 \pm  0.05$ \\
4 & 2000 Dec. 12.82--14.75 & 30  & 80  & --     & 38  &  &  $<9.8 \times 10^{-4}$  &
     $<5.2 \times 10^{-4}$  &   $<6.7 \times 10^{-2}$ \\
\enddata
\tablecomments{The mean rate in 0.1--2 keV, 2--10 keV and 15--200 keV come from 
LECS, MECS and PDS, respectively. \\
The upper limits are given at 95\% confidence level.}
\label{t:log}
\end{deluxetable}

\clearpage

%
% Figure 1
%

\clearpage

% Table 2
%
\begin{deluxetable}{l c c c c c c c c}
\tabletypesize{\small}
\tablewidth{0pt}
%\tablenum{1}
\tablecaption{$\chi^2/{\rm dof}$ (in parenthesis the  null hypothesis probability) for 
different models in  observations 1, 2 and 3.
}
\tablehead{
\colhead{TOO} & \colhead{CPL} & \colhead{COMP,NR} & \colhead{REF/SPH} & \colhead{REF/SL} & 
\colhead{$+$Z/SPH} & \colhead{$+$Z/SL} & \colhead{10eV/SPH} & \colhead{10eV/SL} \\
\colhead{} & \colhead{(1)} & \colhead{(2)} & \colhead{(3)} & \colhead{(4)} & 
\colhead{(5)} & \colhead{(6)} & \colhead{(7)} & \colhead{(8)}
}
\startdata
1 & 222/176 & 223/176 & 203/174 & 211/174 & 195/173 & 194/173 & 192/173 & 192/173 \\
  & (1\%) & (1\%) & (6\%) & (3\%) & (11\%) & (13\%) & (14\%) & (14\%) \\
2 & 126/108  & 126/108  & 102/106  & 108/106  & 96/105   &  95/105  & 98/105   & 95/105   \\
  & (11\%) & (11\%) & (58\%) & (42\%) & (72\%) & (75\%) & (68\%) & (75\%) \\
3 & 212/168  & 212/168  & 183/166  & 193/166  & 165/165  & 161/165  & 167/165  & 163/165  \\
  & (1\%) & (1\%) & (18\%) & (7\%) & (49\%) & (57\%) & (44\%) & (53\%) \\
\enddata
\tablecomments{In the fits $N_{\rm H}$  was free to vary. In TOO2 the LECS was 
switched off and the fit was done from to 2 to 200 keV. 
In this case $N_{\rm H}$, $T_{\rm bb}$ and $L_{\rm bb}$ were fixed to the values derived 
from TOO1. (1) = {\sc bb}+{\sc cutoffpl}; (2) = {\sc bb}+{\sc comptt}, sph, no reflection; 
(3) = {\sc bb}+{\sc compps}, sph, Z/Z$_\odot$=1, reflection; (4) = {\sc bb}+{\sc compps}, sl, 
Z/Z$_\odot$=1, reflection;
(5) = {\sc bb}+{\sc compps}, sph, Z/Z$_\odot$ free, reflection; (6) =  {\sc bb}+{\sc compps}, sl, 
Z/Z$_\odot$ free, reflection; (7) = {\sc bb}+{\sc compps}, sph, Z/Z$_\odot$ free, $kT_{seed} = 10$~eV, 
reflection; 
(8) = {\sc bb}+{\sc compps}, sl, Z/Z$_\odot$ free, $kT_{seed} = 10$~eV, reflection. 
}
\label{t:chi2}
\end{deluxetable}

\clearpage

% Table 3 
%
\begin{deluxetable}{l c c c}
\tabletypesize{\small}
\tablewidth{0pt}
%\tablenum{1}
\tablecaption{Best fit parameters of the {\sc bb+compps} (slab, $Z/Z_{\odot}$ free) in 
the first three observations.}
\tablehead{
\colhead{Parameter} & \colhead{TOO1} & \colhead{TOO2} & \colhead{TOO3} \\
}
\startdata
$\nh\, [10^{20}\,$cm$^{-2}$] & $1.50^{+0.18}_{-0.17}$ &  [1.28] 
                             & $0.74^{+0.25}_{-0.18}$  \\
$kT_{\rm bb}$ [eV] & $35^{+3}_{-4}$  & [35] & $52^{+7}_{-6}$ \\
$L_{\rm bb}\,[10^{34}\,$erg/s]\tablenotemark{a} & $5.8^{+2.2}_{-1.3}$ 
                         & [5.8]                & $1.3^{+2.0}_{-1.2}$ \\
$R_{bb}/R_g$\tablenotemark{a} & $35^{+7}_{-4}$ & -- &  $8^{+13}_{-4}$ \\
$kT_{\rm seed}$ [eV]   &  $= kT_{\rm bb}$  & $= kT_{\rm bb}$   &  $= kT_{\rm bb}$ \\
$kT_{\rm e}$~[keV] &  $63^{+20}_{-11}$ &  $96^{+37}_{-29}$ &  $90^{+12}_{-13}$   \\
$\tau$ &  $0.64^{+0.20}_{-0.11} $ & $0.75^{+0.29}_{-0.24}$ &  $0.80^{+0.11}_{-0.12}$  \\
$\Omega/2 \pi$ & $0.16^{+0.11}_{-0.06}$ &  $0.22^{+0.08}_{-0.12}$  & 
                 $0.24^{+0.08}_{-0.06}$ \\
$Z/Z_\odot$ &  $0.10^{+0.11}_{-0.06}$ & $0.17^{+0.14}_{-0.09}$ & $0.15^{+0.09}_{-0.07}$ \\
$I_{Fe}$ [$10^{-4}$~cm$^{-2}$ s$^{-1}$] & $\le 2.4 $ ($2\sigma$) & 
            $\le 2.1$ ($2\sigma$) & $\le 1.8$ ($2\sigma$)  \\
$EW$ (eV)   &  $\le 33 $ ($2\sigma$) & $\le 31 $ ($2\sigma$) & $\le 30 $ ($2\sigma$) \\
\noalign{\smallskip}

\noalign{\smallskip}
$F_{\rm 0.1--2~keV}$\tablenotemark{b}   & 5.73	     &   --	    &	5.77         \\
$F_{\rm 2--10~keV}$\tablenotemark{b}    & 6.95	     &  6.62	    &   5.89         \\
$_{\rm 15--200~keV}$\tablenotemark{b}   & 19.4	     &  18.2	    &   15.7	     \\
$\chi^2/{\rm dof}$  &  194/173\tablenotemark{c} &  95/105 &  161/165  \\
\enddata
\tablenotetext{a}{Scaled to $d=1.5$ kpc.}
\tablenotetext{b}{in units of $10^{-10}$~erg~cm$^{-2}$~s$^{-1}$.}
\tablenotetext{c}{The  $\chi^2/{\rm dof}$ value is different from that reported in F01, 
due to a further rebinning of the HPGSPC count rate spectrum. With this new binning,
the fit results are unchanged with respect to F01.}  
\label{t:results_1}
\end{deluxetable}

\clearpage

%
% Figure 2
%

%
% Figure 3
%

\clearpage

%
% Figure 4
%

\clearpage

% Table 4
%
\begin{deluxetable}{l c c c}
\tabletypesize{\small}
\tablewidth{0pt}
%\tablenum{1}
\tablecaption{Best fit parameters of the {\sc bb+compps} model, assuming a slab geometry, 
$Z/Z_{\odot}$ free and $kT_{\rm seed} = 10$~eV.}
\tablehead{
\colhead{Parameter} & \colhead{TOO1} & \colhead{TOO2} & \colhead{TOO3} \\
}
\startdata
$\nh\, [10^{20}\,$cm$^{-2}$] & $1.73^{+0.33}_{-0.35}$ &  [1.28] 
                             & $1.09^{+0.29}_{-0.21}$  \\
$kT_{\rm bb}$ [eV] &  $32^{+4}_{-2}$   & -- & $43^{+7}_{-5}$ \\
$L_{\rm bb}\,[10^{34}\,$erg/s]\tablenotemark{a} & $13^{+12}_{-8}$ 
                         & --                & $6.7^{+4.5}_{-2.7}$ \\
$R_{bb}/R_g$\tablenotemark{a} & $67^{+34}_{-22}$ & -- & $26^{+12}_{-8}$  \\
$kT_{\rm seed}$ [eV]   & [10]  & [10]   &  [10] \\
$kT_{\rm e}$~[keV] &  $68^{+17}_{-18}$ &  $91^{+79}_{-26}$ &  $112^{+23}_{-30}$   \\
$\tau$ &  $0.59^{+0.15}_{-0.15} $ & $0.80^{+0.60}_{-0.23}$ &  $0.61^{+0.13}_{-0.17}$  \\
$\Omega/2 \pi$ & $0.18^{+0.12}_{-0.09}$ &  $0.21^{+0.12}_{-0.11}$  & 
                 $0.27^{+0.08}_{-0.09}$ \\
$Z/Z_\odot$ &  $0.10^{+0.10}_{-0.06}$ & $0.16^{+0.13}_{-0.09}$ & $0.14^{+0.10}_{-0.06}$ \\
$I_{Fe}$ [$10^{-4}$~cm$^{-2}$ s$^{-1}$] & $\le 2.4 $ ($2\sigma$) & 
            $\le 2.1$ ($2\sigma$) & $\le 1.8$ ($2\sigma$)  \\
$EW$ (eV)    &  $\le 33 $ ($2\sigma$) & $\le 31 $ ($2\sigma$) &  $\le 30 $ ($2\sigma$) \\ 
$\chi^2/{\rm dof}$  &  192/173\tablenotemark{b} &  95/105 &  163/165  \\
\enddata
\tablecomments{sl = slab}
\tablenotetext{a}{Scaled to $d=1.5$ kpc.}
\tablenotetext{b}{See corresponding note in Table~\ref{t:results_1}.} 
\label{t:results_2}
\end{deluxetable}

\clearpage

% Table 5 
%
\begin{deluxetable}{l c c c c c c c}
\tabletypesize{\small}
\tablewidth{0pt}
%\tablenum{1}
\tablecaption{Fitting results of the 0.01--50 Hz power spectral density}
\tablehead{
\colhead{TOO} & \colhead{Energy band} & \colhead{$\beta$} & \colhead{$R_t$} & 
 \colhead{$f_{QPO}$} & 
            \colhead{$\Delta_{QPO}$} & \colhead{$R_{QPO}$}  & \colhead{$R_{QPO}/R_t$} \\
       \colhead{} & \colhead{(keV)} & \colhead{} & \colhead{\% ($rms$)} & \colhead{($\times 
10^{-2}$~Hz)} & 
       \colhead{($\times 10^{-2}$~Hz)} & \colhead{\% ($rms$)} & \colhead{ } 
}
\startdata
 1 & 0.1--0.5   &   1.13$_{-0.06}^{+0.05}$  & 67$\pm$7  &  [7.81]    &             
       2.0$_{-1.0}^{+1.3}$   &  $< 12.8$ (2$\sigma$)   &  $< 0.19$ (2$\sigma$)  \\
   &  0.1--1.5  &   1.30$_{-0.10}^{+0.09}$  & 62$\pm$3  &  7.81$_{-0.25}^{+0.29}$ & 
   0.96$_{-0.59}^{+1.04}$ &  $8.1_{-1.4}^{+1.5}$ &  0.13$_{-0.02}^{+0.03}$     \\
   & 1.6--10     &  $1.15_{-0.03}^{+0.03}$  &   42$\pm$2  & 8.11$_{-0.17}^{+0.16}$ & 
   1.23$_{-0.42}^{+0.57}$ & $9.2_{-0.9}^{+0.7}$  &  $0.22_{-0.02}^{+0.01}$       \\
   & 15-200     &   1.25$_{-0.35}^{+0.10}$ &  18$\pm$1   &  7.82$_{-0.29}^{+0.39}$ & 
   2.40$_{-1.37}^{+9.89}$ &  $5.4_{-1.0}^{+0.6}$  &  $0.30_{-0.05}^{+0.03}$     \\ 
   &           &           &              &             &            &           \\
 2 & 0.1--0.5  &   --   &   --    &   --   &  --   &  --   \\     
   & 0.1--1.5  & --   &   --    &   --   &  --   &  --   \\  
   & 1.6--10   & 1.26$_{-0.04}^{+0.04}$   &    38$\pm$4  &  12.08$_{-0.11}^{+0.12}$  &  
   1.25$_{-0.27}^{+0.36}$  &  $9.9_{-0.7}^{+0.7}$  &   $0.26_{-0.02}^{+0.02}$    \\
   & 15--200   & 1.39$_{-0.16}^{+0.13}$   &    16$\pm$1  &   12.20$_{-0.41}^{+0.35}$ &
   1.89$_{-0.80}^{+1.30}$  &  $3.8_{-0.6}^{+0.4}$  &   $0.24_{-0.04}^{+0.02}$    \\
   &           &           &              &             &            &           \\
 3 & 0.1--0.5  &  1.71$_{-0.35}^{+0.26}$  &   56$\pm$8   &    [7.29]               &
   2.82$_{-1.15}^{+1.65}$  &  19.1$_{-4.1}^{+3.2}$ &    $0.34_{-0.07}^{+0.06}$   \\  
   & 0.1--1.5  &  1.07$_{-0.08}^{+0.09}$  &   52$\pm$4   &    7.29$_{-0.33}^{+0.41}$  &
   1.49$_{-0.63}^{+0.98}$   &  10.7$_{-1.2}^{+1.0}$ &    $0.21_{-0.03}^{+0.02}$    \\
   & 1.6--10   &  1.21$_{-0.04}^{+0.03}$  &   32$\pm$2   &    [7.29]                  &
   0.99$_{-0.51}^{+8.10}$  &  $< 5.9$ (2$\sigma$)  &            $< 0.19$(2$\sigma$) \\
   & 15--200   &  1.31$_{-0.65}^{+0.23}$  &   14$\pm$1   &    [7.29]                  &
   1.05$_{-0.59}^{+1.09}$  &    $< 3.0$ (2$\sigma$) &    $< 0.21$ (2$\sigma$)  
\enddata
\label{t:time}
\end{deluxetable}

\clearpage

%
% Figure 5
%

\clearpage

%
% Figure 6
%

\clearpage

%
% Figure 7
%

\clearpage
%
%
% Figure 8
%

\end{document}